\documentclass[12pt,preprint]{aastex}

\slugcomment{Submitted to ApJ}

\shorttitle{Effects of Rotation on Supernova Explosions}
\shortauthors{YAMASAKI \& YAMADA}

\begin{document}

\title{Effects of Rotation on the Revival of \\
a Stalled Shock in Supernova Explosions}

\author{Tatsuya Yamasaki\altaffilmark{1} and Shoichi Yamada\altaffilmark{2,3}}

\altaffiltext{1}{Department of Astronomy, Faculty of Science,
Kyoto University, Oiwake-cho, Sakyo, Kyoto 606-8502, Japan;
yamasaki@kusastro.kyoto-u.ac.jp}

\altaffiltext{2}{Science and Engineering, Waseda University, 3-4-1 Okubo,
Shinjuku, Tokyo 169-8555, Japan;
shoichi@heap.phys.waseda.ac.jp}

\altaffiltext{3}{Advanced Research Institute for Science and Engineering,
Waseda University, 3-4-1 Okubo, Shinjuku, Tokyo 169-8555, Japan}

\begin{abstract}
In order to infer the effects of rotation on the revival of a stalled shock
in supernova explosions, we investigated steady 
accretion flows with a standing shock.
We first obtained a series of solutions for equations describing non-rotating
spherically symmetric flows and confirmed the results of preceding papers that, 
for a given mass accretion rate, there is a critical luminosity of irradiating 
neutrinos, above which there exists no steady solution.
Below the critical value, we found two branches of solutions;
one is stable and the other is unstable against radial perturbations.
With a simple argument based on the Riemann problem, 
we can identify the critical luminosity as the one,
at which the stalled shock revives.
We also obtained the condition satisfied by the flow velocity
for the critical luminosity, which can be easily applied to the rotational case.
If a collapsing star rotates, the accretion flow is non-spherical due to 
centrifugal forces.
Flows are accelerated near the rotation axis whereas they 
are decelerated near the equatorial plane.
As a result, the critical luminosity is lowered,
that is, rotation assists the revival of a stalled shock.
According to our calculations, the critical luminosity is $\sim25$\% lower 
for the mass accretion rate of 1M$_{\odot}$/sec and the rotation frequency
of 0.1 Hz at a radius of 1000 km than that of the spherically symmetric flow
with the same mass accretion rate.
We found that the condition of the flow velocity 
at the critical luminosity is first satisfied at the rotation axis.
This suggests that the shock revival is triggered on the rotation axis and 
a jet-like explosion ensues.
\end{abstract}

\keywords{stars: rotation ---
shock waves ---
hydrodynamics ---
supernovae: general}

\section{Introduction}

Core collapse supernovae play important roles, for example, in star formations,
evolutions of galaxy and the entire universe,
and accelerations of cosmic ray particles
owing to its nucleosynthesis, energetic shock waves and high luminosity.
Recently the phenomena are put in the spotlight
because the gamma-ray bursts are likely to be associated with them.
Unfortunately, however, the mechanism of supernova explosion is
still unresolved [e.g. \citet{whe03} for reviews].
 
The supernova explosions are commenced by gravitational collapse of
massive stars at the end of their lives.
When the nuclear saturation density is reached at the center of core,
the bounce occurs, producing a shock wave that starts to 
propagate outwards in the core.
It is almost a consensus among the researchers that 
this shock wave does not yield the supernova explosion directly,
since the shock loses energy by photo-dissociations of nuclei
as well as neutrino-cooling and stalls somewhere inside the core.
It is widely expected that the shock will be revived
by the irradiation of neutrinos diffusing out of the proto-neutron star,
the scenario originally proposed by \citet{wil82}. 
Although many researchers have studied this scenario
intensively and extensively, the state-of-the-art numerical simulations
have shown so far that, as long as collapse is spherically symmetric,
the stalled shocks do not revive \citep{lie01,bur03,tho03,lie04}.

These simulations are concerned with the dynamical evolution
of the entire core from the onset of the collapse till long after
the stagnation of the shock. 
On the contrary, \citet{bur93} took a different approach,
focusing on the revival of the stalled shock wave.
After the shock is stagnated due to the energy losses mentioned above,
the accretion flows through the standing shock wave are quasi-steady
(see \citet{jan01} for the criticism on this point).
They approximated these flows with steady state solutions
for constant mass accretion rates and neutrino luminosities.
Varying these constant values, they found that for a given mass accretion rate 
there is a critical neutrino luminosity,
above which there exists no steady solutions.
Using this fact, they argued that the revival of stalled shock occurs when 
the neutrino luminosity exceeds this critical value.
The existence of the critical luminosity was also observed
in the numerical experiments done by \citet{jan96},
where they artificially increased neutrino luminosities
in their failed explosion models and saw what happened to the stalled shocks.
Merits of these steady state calculations are not only that
they are computationally simpler but also that they can clearly demonstrate,
if qualitatively, what is the cause of the failure of explosions. That 
is, according to this theory, an inappropriate combination of the neutrino 
luminosity and the mass accretion rate.

So far, we have been talking about the studies of
spherically symmetric collapse.
However, several observations suggest that the supernova explosions are
intrinsically asymmetric in general. For example, 
\citet{leo00} and \citet{wan02} observed a few percent of
linear polarizations for photons from collapse-driven supernovae
and argued that the stellar envelopes are globally asymmetric,
oblate or prolate with an aspect ration of $\sim 2$.
We also know that supernova remnants are asymmetric in general. 
We can easily imagine a couple of possible causes for this asymmetry, e.g. 
hydrodynamical instabilities, rotation and magnetic fields.
In fact, young pulsars are thought to have rather large angular momenta
\citep{kas94,swa01} or magnetic fields \citep{kul92}, which are probably taken 
over from the pre-collapse stars.
It is also believed that some convective motions occur in the supernova core.
The large scale non-spherical oscillations of the standing accretion shocks
may also cause the stellar envelope globally asymmetric\citep{blo03}. 

Among these effects, we pay particular attention to the 
stellar rotation in this paper.
Note, however, that the rotation of the core may not be so rapid after all.
Recent evolution models by Heger et al.~\citep{heg00,heg03} suggest that
the transport of angular momentum during the quasi-static evolutionary 
phase of the progenitor deprives the core of substantial fraction of
its angular momentum, particularly when the magnetic torque is taken into 
account~\citep{spr02,heg03}.
If this is really the case, the rotation will play no 
significant role in dynamics of core-collapse as shown by~\citet{bur03} 
(see also~\citet{mue04}). We had better bear in mind, however, 
that the evolution models are based on 1D calculations and have
some uncertainties in the mechanism and treatment of angular momentum transport.

The dynamics of rotational collapse has been explored numerically
by some authors.
For example, \citet{moe89} and \citet{yam94} demonstrated that rapid rotation
tends to weaken the prompt shock wave because the centrifugal force prevents
the core from contracting sufficiently.
\citet{fry00} found that the efficiency of energy transfer by convection was
reduced by the decrease of effective gravity near the equatorial plane due to 
the centrifugal force and, as a result, the shock did not propagate as much 
as in the non-rotation case.
They also asserted that the resultant anisotropy of energy transfer would
have made explosion jet-like if it had been successful.
\citet{kot03} estimated the anisotropy of neutrino radiation from the 
rotational cores. They claimed that the deformed neutrino-sphere,
oblate in general, yields the neutrino flux and the heating rate
that are larger near the rotation axis than near the equatorial plane.
Given the results by \citet{kot03} and the results by \citet{shi01} that
such anisotropic neutrino irradiation could lead to successful jet-like
explosion, we may well expect that the rotation is helpful for the explosion
in the neutrino heating scenario.

In order to see the effect of rotation more clearly, if qualitatively,
we followed the lead by \citet{bur93}.
We solved time-independent Euler equations describing steady accretion flows
through a standing shock, taking rotation into account.
We paid particular attention to how the critical luminosity of neutrinos
is altered by the stellar rotation and demonstrated that the rotation assist
the revival of the stalled shock.

This paper is organized as follows.
In the next section, we discuss the spherically symmetric case.
We find two branches of steady solutions and discuss the revival of 
the shock wave employing the critical luminosity.
We examine in section 3 the stability of the two branches of solutions
obtained in section 2.
In section 4, we analyze the steady axisymmetric solutions with rotation and 
discuss the effect of rotation on the revival of the stalled shock.
The final section is devoted to summary and some more discussion.

\section{Spherically Symmetric Flows With a Standing Shock}

In this and next sections,
we discuss spherically symmetric flows, solving one dimensional
hydrodynamical equations.
The calculations in this section are similar to those of \citet{bur93}.
After the shock is stalled, mass accretion rates
or neutrino luminosities change
rather slowly and the flows become quasi-steady.
Idealizing this situation,
we seek in this section steady solutions of the accretion flow.

Since our purpose of this paper is not to determine quantitatively the critical
luminosity of neutrinos, we make following assumptions and simplifications:
i) We do not consider the flows outside the shock wave in detail,
simply assuming a free fall, and restrict
our calculations to the inside of the shock surface.
ii) Newtonian formulations are adopted because the region of our interest
is outside of the neutron star
and general relativistic effects are negligible there.
iii) We do not solve neutrino transfer equations assuming that the luminosity
and energy of the neutrinos are independent of radius. 
It is also assumed that the neutrinos are thermal
with a fixed value of temperature ($T_{\nu}=4.5 {\rm MeV}$), for simplicity.
iv) A simplified equation of state and rates of neutrino heating
and cooling are employed.
v) Photo dissociations of nuclei,
magnetic fields and convections are neglected, for simplicity,
although there should exist regions where flows are convectively unstable.

Then, the basic equations describing steady spherically symmetric
accretion flows are given in the spherical coordinate as,

\begin{equation}
4\pi r^2 \rho u_r=\dot{M},
\label{eq1}
\end{equation}

\begin{equation}
u_r \frac{du_r}{dr}+\frac{1}{\rho}\frac{dp}{dr}+\frac{GM}{r^2}=0,
\label{eq2}
\end{equation}

\begin{equation}
u_r \frac{d\epsilon}{dr}-\frac{p}{\rho^2}u_r \frac{d\rho}{dr}
=\dot{q},
\label{eq3}
\end{equation}
where $u_r$, $\rho$, $p$, $\epsilon$ denote radial velocity, density,
pressure and specific internal energy, respectively;
$\dot{M}$, $G$, $M$, $\dot{q}$ are mass accretion rate,
gravitational constant, neutron star mass
and net heating rate by neutrino irradiation, respectively.
As for the neutron star mass, we adopt take $1.3M_{\odot}$.

Pressure and specific internal energy of nucleons, photons,
and relativistic particles are approximately written as

\begin{equation}
p=\frac{11\pi^2}{180}\frac{k^4}{c^3 \hbar^3}T^4 +\frac{\rho kT}{m_{\rm N}},
\label{eq4}
\end{equation}

\begin{equation}
\epsilon=\frac{11\pi^2}{60}\frac{k^4}{c^3 \hbar^3}\frac{T^4}{\rho}
+\frac{3}{2}\frac{kT}{m_{\rm N}},
\label{eq5}
\end{equation}
where $k$, $c$, $\hbar$, $m_{\rm N}$ are Boltzmann constant,
speed of light, Planck constant, and nucleon mass, respectively;
$T$ is temperature of matter and photons.

As for the heating and cooling processes,
we take into account only the absorption and emission of neutrinos
by nucleons.
We adopt the formulae given in \citet{her92} for degeneration factor
$\eta=0$; that is

\begin{equation}
\dot{q}=4.8\times 10^{32}\frac{L_{\nu_{\rm e}}({\rm foes}~{\rm s}^{-1})}
{4\pi r^2}T_{\nu}^2
-2.0\times 10^{18}T^{6}~({\rm ergs}~{\rm s}^{-1} {\rm g}^{-1}),
\label{eq6}
\end{equation}
where $L_{\nu_{\rm e}}$ is the luminosity of electron neutrinos,
which is assumed to be equal to that of anti-electron neutrinos.

The outer and inner boundaries of calculations are
set at the shock surface and the neutrino-sphere, respectively.
We impose the Rankine-Hugoniot relations at the shock surface.
The flow into the shock is assumed to be a free fall
with the velocity of $u_{\rm f}=-\sqrt{2GM/r}$ and the temperature of $0$.
Thus, the outer boundary conditions are written as

\begin{equation}
\rho u_r^2 +p=\rho_{\rm f}u_{\rm f}^2,
\label{eq7}
\end{equation}

\begin{equation}
\frac{1}{2} u_r^2 +\epsilon +\frac{p}{\rho}=\frac{1}{2} u_{\rm f}^2,
\label{eq8}
\end{equation}
where $\rho_{\rm f}$ is upstream density at the shock surface.
The continuity equation is automatically satisfied by imposing
equation (\ref{eq1}) on each side of the shock.
We impose the condition that the density be $10^{11}$ ${\rm g/cm^3}$
at the inner boundary; this is approximately corresponding to the condition 
that the optical depth from the neutrino-sphere to infinity equals to $2/3$.
The latter condition was adopted by \citet{bur93}.

We note that the position of the inner boundary is determined by the relation

\begin{equation}
L_{\nu_{\rm e}}=\frac{7}{16}\sigma T_{\nu}^4 \cdot 4\pi r_{\nu}^2 ,
\label{eq9}
\end{equation}
where $\sigma$ is the Stefan-Boltzmann constant and
$r_{\nu}$ is the radius of the neutrino-sphere,
that is, the inner boundary.
On the other hand, the position of the outer boundary is
determined so as to satisfy the above jump conditions.
Thus, the problem is regarded as an eigen-value problem.
We solve these equations for a wide range of values
of the mass accretion rate and the neutrino luminosity.

\begin{figure}
\plottwo{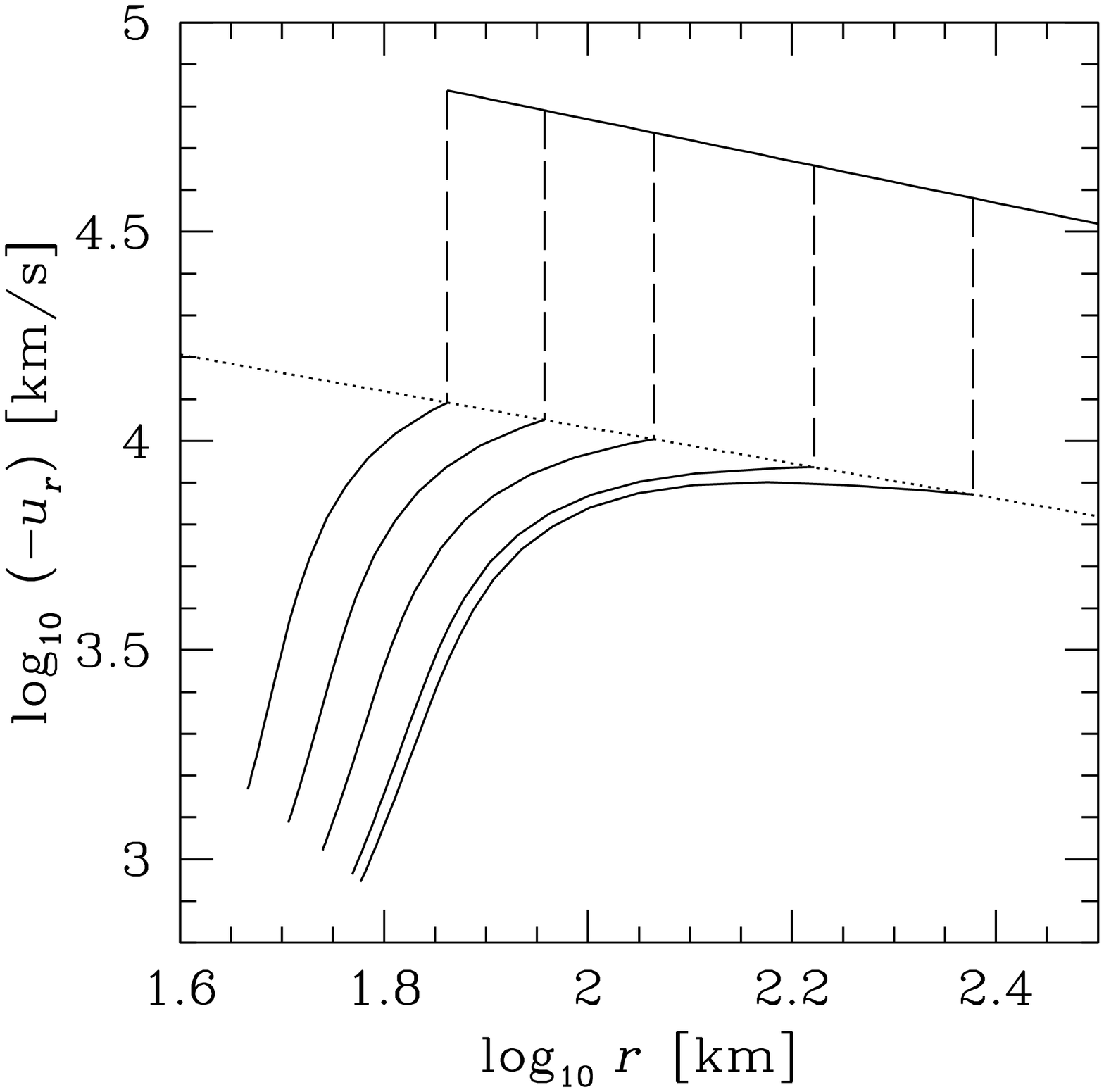}{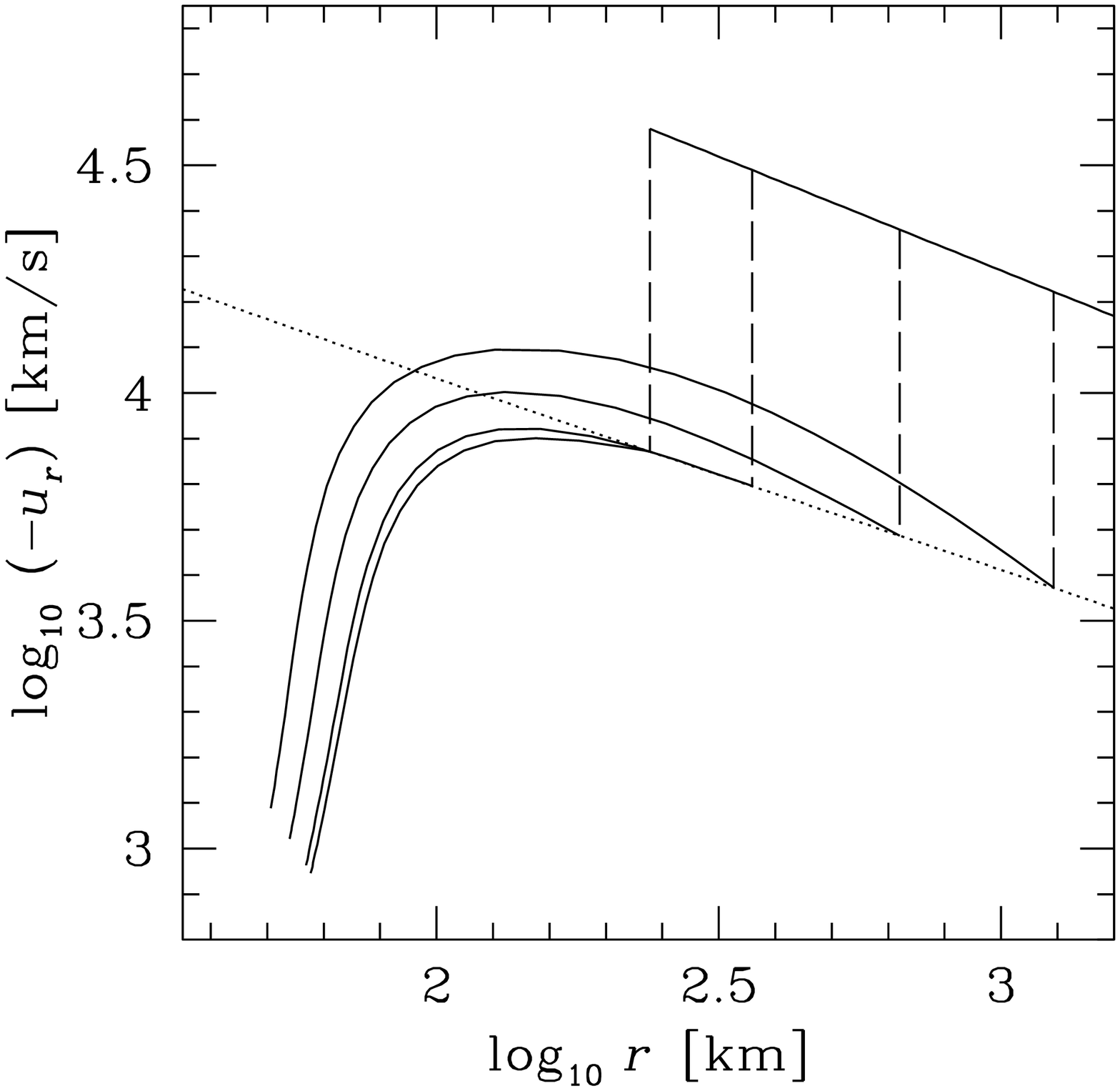}
\caption{a. Inner solution curves for spherical steady accretion flows
with $\dot{\rm M}=2.0$ ${\rm M}_{\odot}/{\rm s}$,
${\rm L}_{\nu_{\rm e}}=5,6,7,8\cdot 10^{52}$ ergs/s
and $8.3167 \cdot 10^{52}$ ergs/s
(critical value) from left to right.
b. Outer solution curves for steady accretion flows
with $\dot{\rm M}=0.2$ ${\rm M}_{\odot}/{\rm s}$,
${\rm L}_{\nu_{\rm e}}=6,7,8\cdot 10^{52}$ ergs/s
and $8.3167 \cdot 10^{52}$ ergs/s from left to right.
Dashed lines denote shock jumps.
Dotted lines show the downstream values
satisfying the Rankine-Hugoniot relations at each radius.
\label{fig1}}
\end{figure}

The results of the calculations are shown in figures \ref{fig1} and \ref{fig2}.
In figure \ref{fig1}, we can see there are two solutions for a given mass
accretion rate and a neutrino luminosity
when the neutrino luminosity is below a certain critical value.
The shock radii of these two solutions differ.
From now on, we refer to the solution with a smaller shock radius
as the inner solution and to the other solution as the outer solution.
As shown in figure \ref{fig2},
as the neutrino luminosity is raised with the mass accretion rate fixed,
the shock radius for the inner solutions becomes larger,
whereas that for the outer solutions becomes smaller.
Two solutions coincide with each other
when the luminosity reaches the critical value.
For the luminosity over the critical value, there is no solution.
This critical value of neutrino luminosity is shown in figure \ref{fig8}.
Note that when the luminosity is close to the critical value,
the shock radii are insensitive to the luminosity.
Although our formulation is different in detail from Burrows \& Goshys',
the results are in good agreement [cf. figure 1 of \citet{bur93}].
As mentioned, the shock radius for the critical luminosity is
almost independent of the mass accretion rate,
and is about $200-300$ km according to our calculations.

This behavior of the solutions is understood if one remind of
the adiabatic subsonic Bondi-Hoyle flow (cf. \citet{sha83}).
Here, we introduce $u_{\rm post}(r)$, as the post shock velocity
that we would obtain if the shock radius were $r$.
The velocity $u_{\rm post}(r)$ is shown as dotted lines in figure \ref{fig1}.
The subsonic adiabatic flows accelerate outside a certain radius,
whereas they decelerate inside that radius.
The critical radius is determined by
the balance of the gravitational energy and thermal energy
and is close to the sonic point in the transonic accretion flow.
On the other hand, the free fall velocity decreases monotonically.
Furthermore, the acceleration of the adiabatic flow is larger than
that of the free fall outside the critical radius
owing to the pressure gradient force.
Therefore, if any,
there are two radii where the accretion flow joins
with the free fall stream via the Rankine-Hugoniot relations.
In other words, the solution curve of the adiabatic accretion flow $u_r(r)$
intersects with $u_{\rm post}(r)$ at two points.
Although the solution curves are somewhat modified if one takes account of
the neutrino irradiation, this behavior is not altered,
for the amplitudes of neutrino luminosity of our interest.
The irradiation tends to heat and push out the infalling matter.
Hence, as the luminosity gets larger, the infall velocity gets smaller
in the whole range and the solution curves go down.
When the luminosity exceeds a critical value,
there exists no point where the subsonic accretion flow
joins with the free fall stream, via the Rankine-Hugoniot relations;
the solution curves of the accretion flow $u_r(r)$
never intersects with $u_{\rm post}(r)$.

From the above explanation,
we can also understand another important nature of the solutions;
as the luminosity is raised with the mass accretion rate fixed,
the radial derivative of the velocity of the accretion flow at the shock
radius $du_r(r)/dr$ becomes larger for the inner solutions,
whereas it becomes smaller for the outer solutions.
If the flow is adiabatic,
these derivatives coincide with that of
$du_{\rm post}(r)/dr$ at the shock radius for the critical luminosity.
The neutrino heating, however, makes
the former slightly smaller than the latter.
This characteristics will be utilized in section 4.

Finally, let us discuss what happens when the luminosity exceeds the
critical value.
Since there is no steady solution, we should discuss this issue
with time-dependent calculations.
However, we can infer at least the initial response of the shock
with a simpler discussion.
We can find in figure \ref{fig1} that
when the luminosity exceeds the critical value,
the downstream velocity at the shock wave becomes
inevitably smaller than $u_{\rm post}(r)$.
In addition, the downstream entropy
should be larger than that for the critical luminosity,
because of the higher luminosity.
We can regard this situation as a Riemann problem
that has a discontinuity
with a velocity slightly lower and an entropy slightly
larger in the downstream than those satisfying
the Rankine-Hugoniot relations.
The solution of the Riemann problem tells us that,
a shock wave will propagate outwards,
while a weak rarefaction wave or another weak shock wave will propagate
inwards at the same time \citep{cou76}; in our words, the shock wave revives.
Thus, we can conclude that the condition for the shock to revive
is that the neutrino luminosity exceeds the critical value.

\section{Stability against Radial Perturbations}

In the previous section, we found that there are two branches of
steady solutions for the accretion flow with a shock.
In this section, we discuss which type of the solutions is realized,
examining the stability of these solutions by global linear analysis.
We consider the perturbations of neutrino luminosity,
that is, non-adiabatic effects are taken into account.

In this paper, only the stability
against radial perturbations is considered,
This is because we did not take account of the effect
of convection in obtaining the unperturbed flows in the previous section.
There is, in general, a region where heating prevails over cooling
under the irradiation of neutrinos;
In this region, entropy increases in the direction of gravity
and the flow is convectively unstable.
Thus our solutions are convectively unstable against non-radial
perturbations, which fact is of no interest to us here, though.

Since we consider the perturbations accompanying variations of the
positions of inner and outer boundaries, it is convenient to introduce
a new variable $x$, defined as,

\begin{equation}
x=\frac{r-r_{\nu}}{r_{\rm s}-r_{\nu}},
\label{eq10}
\end{equation}
where $r_{\rm s}$ denotes the shock radius.
Then, the basic equations describing the time-dependent flows
are written as,

\begin{equation}
\frac{\partial \rho}{\partial t}
-\frac{1}{r_{\rm s}-r_{\nu}}
\left\{x\frac{\partial r_{\rm s}}{\partial t}
+(1-x)\frac{\partial r_{\nu}}{\partial t}\right\}
\frac{\partial \rho}{\partial x}
+\frac{2\rho u_r}{r}
+\frac{\rho}{r_{\rm s}-r_{\nu}}
\frac{\partial u_r}{\partial x}
+\frac{u_r}{r_{\rm s}-r_{\nu}}
\frac{\partial \rho}{\partial x}
=0,
\label{eq11}
\end{equation}

\begin{equation}
\frac{\partial u_r}{\partial t}
-\frac{1}{r_{\rm s}-r_{\nu}}
\left\{x\frac{\partial r_{\rm s}}{\partial t}
+(1-x)\frac{\partial r_{\nu}}{\partial t}\right\}
\frac{\partial u_r}{\partial x}
+\frac{u_r}{r_{\rm s}-r_{\nu}}
\frac{\partial u_r}{\partial x}
+\frac{1}{r_{\rm s}-r_{\nu}}\frac{1}{\rho}\frac{\partial p}{\partial x}
+\frac{GM}{r^2}=0,
\label{eq12}
\end{equation}

\begin{eqnarray}
\lefteqn{\frac{\partial \epsilon}{\partial t}
-\frac{1}{r_{\rm s}-r_{\nu}}
\left\{x\frac{\partial r_{\rm s}}{\partial t}
+(1-x)\frac{\partial r_{\nu}}{\partial t}\right\}
\frac{\partial \epsilon}{\partial x}
-\frac{p}{\rho^2}\frac{\partial \rho}{\partial t}} \nonumber\\
& &-\frac{1}{r_{\rm s}-r_{\nu}}
\left\{x\frac{\partial r_{\rm s}}{\partial t}
+(1-x)\frac{\partial r_{\nu}}{\partial t}\right\}
\frac{p}{\rho^2}\frac{\partial \rho}{\partial x}
+\frac{u_r}{r_{\rm s}-r_{\nu}} \frac{\partial \epsilon}{\partial x}
-\frac{u_r}{r_{\rm s}-r_{\nu}}\frac{p}{\rho^2}
\frac{\partial \rho}{\partial x}
-\dot{q}=0.
\label{eq13}
\end{eqnarray}

Since we assume that the flow outside the shock is steady,
the Rankine-Hugoniot relations at the outer boundary are expressed as,

\begin{equation}
\rho(u_r -u_{\rm s})=\rho_{\rm f}(u_{\rm f}-u_{\rm s}),
\label{eq14}
\end{equation}

\begin{equation}
\rho(u_r -u_{\rm s})^2 +p=\rho_{\rm f}(u_{\rm f}-u_{\rm s})^2,
\label{eq15}
\end{equation}

\begin{equation}
\frac{1}{2}(u_r -u_{\rm s})^2 +\epsilon+\frac{p}{\rho}
=\frac{1}{2}(u_{\rm f}-u_{\rm s})^2,
\label{eq16}
\end{equation}
where $u_{\rm s}$ is the shock velocity, which can be written as

\begin{equation}
u_{\rm s}=\frac{\partial r_{\rm s}}{\partial t}.
\label{eq17}
\end{equation}

In calculating the perturbations of heating and cooling rates,
we assume, for simplicity, that the effective temperature
of neutrinos is not perturbed.

We perform a linear analysis, assuming that all the perturbed quantities
have the time-dependence of $\exp(\omega t)$.
In the rest of this section, the suffix $1$ is attached
to the perturbed quantities and the suffix $0$ to the unperturbed ones.
Then, the equations describing the perturbations become

\begin{eqnarray}
\lefteqn{\frac{\partial (\rho_1/\rho_0)}{\partial x}
+\frac{\partial (u_{r1}/u_{r0})}{\partial x}
+\frac{r_{{\rm s}0}-r_{{\nu}0}}{u_{r0}}\omega\frac{\rho_1}{\rho_0}
+\frac{2}{r_0}\left(r_{{\rm s}0}\frac{r_{{\rm s}1}}{r_{{\rm s}0}}
-r_{{\nu}0}\frac{r_{{\nu}1}}{r_{{\nu}0}}\right)} \nonumber\\
& &+\left(2\frac{r_{{\rm s}0}-r_{{\nu}0}}{r_0}
+\frac{\partial \ln \rho_0}{\partial x}
+\frac{\partial \ln u_{r0}}{\partial x}\right)
\left(\frac{\rho_1}{\rho_0}+\frac{u_{r1}}{u_{r0}}\right) \nonumber\\
& &+\left(\frac{1}{r_0}\frac{\partial \ln u_{r0}}{\partial x}
+\frac{1}{r_0}\frac{\partial \ln \rho_0}{\partial x}
-\frac{\omega}{u_{r0}}\frac{\partial \ln \rho_0}{\partial x}\right)
\left\{xr_{{\rm s}0}\frac{r_{{\rm s}1}}{r_{{\rm s}0}}
+(1-x)r_{\nu 0}\frac{r_{\nu 1}}{r_{\nu 0}} \right\}=0,
\label{eq18}
\end{eqnarray}

\begin{eqnarray}
\lefteqn{\frac{\partial (u_{r1}/u_{r0})}{\partial x}
+\frac{1}{u_{r0}^2}\frac{p_{{\rm N}0}}{\rho_0}
\frac{\partial (\rho_1/\rho_0)}{\partial x}
+\frac{1}{u_{r0}^2}\frac{4p_{{\rm R}0}+p_{{\rm N}0}}{\rho_0}
\frac{\partial (T_1/T_0)}{\partial x}} \nonumber\\
& &+\left(\frac{r_{{\rm s}0}-r_{{\nu}0}}{u_{r0}}\omega
+2\frac{\partial \ln u_{r0}}{\partial x}\right)
\frac{u_{r1}}{u_{r0}}
-\frac{1}{u_{r0}^2}\frac{4p_{{\rm R}0}}{\rho_0}
\frac{\partial \ln T_0}{\partial x}\frac{\rho_1}{\rho_0} \nonumber\\
& &+\left(\frac{1}{u_{r0}^2}\frac{p_{{\rm N}0}}{\rho_0}
\frac{\partial \ln \rho_0}{\partial x}
+\frac{1}{u_{r0}^2}\frac{16p_{{\rm R}0}+p_{{\rm N}0}}{\rho_0}
\frac{\partial \ln T_0}{\partial x}\right)\frac{T_1}{T_0} \nonumber\\
& &+\left\{\left(\frac{2}{r_0}-\frac{\omega}{u_{r0}}\right)
\frac{\partial \ln u_{r0}}{\partial x}
+\frac{2}{r_0}\frac{1}{u_{r0}^2}\frac{p_{{\rm N}0}}{\rho_0}
\frac{\partial \ln \rho_0}{\partial x}
+\frac{2}{r_0}\frac{1}{u_{r0}^2}\frac{4p_{{\rm R}0}+p_{{\rm N}0}}{\rho_0}
\frac{\partial \ln T_0}{\partial x}\right\} \nonumber\\
& &~~~~~~~~~\times\left\{xr_{{\rm s}0}\frac{r_{{\rm s}1}}{r_{{\rm s}0}}
+(1-x)r_{\nu 0}\frac{r_{\nu 1}}{r_{\nu 0}} \right\} \nonumber\\
& &+\frac{GM}{r_0^2 u_{r0}^2}
\left(r_{{\rm s}0}\frac{r_{{\rm s}1}}{r_{{\rm s}0}}
-r_{{\nu}0}\frac{r_{{\nu}1}}{r_{{\nu}0}}\right)=0,
\label{eq19}
\end{eqnarray}

\begin{eqnarray}
\lefteqn{-\frac{4p_{{\rm R}0}+p_{{\rm N}0}}{\rho_0}
\frac{\partial (\rho_1/\rho_0)}{\partial x}
+\frac{12p_{{\rm R}0}+(3/2)p_{{\rm N}0}}{\rho_0}
\frac{\partial (T_1/T_0)}{\partial x}} \nonumber\\
& &+\left\{-\frac{4p_{{\rm R}0}+p_{{\rm N}0}}{\rho_0}
\frac{\partial \ln \rho_0}{\partial x}
+\frac{12p_{{\rm R}0}+(3/2)p_{{\rm N}0}}{\rho_0}
\frac{\partial \ln T_0}{\partial x}\right\}\frac{u_{r1}}{u_{r0}} \nonumber\\
& &+\left\{-(r_{{\rm s}0}-r_{{\nu}0})\frac{4p_{{\rm R}0}+p_{{\rm N}0}}{\rho_0}
\frac{\omega}{u_{r0}}
+\frac{4p_{{\rm R}0}}{\rho_0}\frac{\partial \ln \rho_0}{\partial x}
-\frac{12p_{{\rm R}0}}{\rho_0}\frac{\partial \ln T_0}{\partial x}
-\frac{r_{{\rm s}0}-r_{{\nu}0}}{u_{r0}}
\frac{\partial \dot{q}}{\partial \rho}\rho_0 \right\}
\frac{\rho_1}{\rho_0} \nonumber\\
& &+\left\{(r_{{\rm s}0}-r_{{\nu}0})
\frac{12p_{{\rm R}0}+(3/2)p_{{\rm N}0}}{\rho_0}\frac{\omega}{u_{r0}}
-\frac{16p_{{\rm R}0}+p_{{\rm N}0}}{\rho_0}
\frac{\partial \ln \rho_0}{\partial x} \right. \nonumber\\
& &~~~~~~~~~+\left.\frac{48p_{{\rm R}0}+(3/2)p_{{\rm N}0}}{\rho_0}
\frac{\partial \ln T_0}{\partial x}
-\frac{r_{{\rm s}0}-r_{{\nu}0}}{u_{r0}}
\frac{\partial \dot{q}}{\partial T}T_0 \right\}
\frac{T_1}{T_0} \nonumber\\
& &+\left\{\frac{4p_{{\rm R}0}+p_{{\rm N}0}}{\rho_0}
\frac{\omega}{u_{r0}}\frac{\partial \ln \rho_0}{\partial x}
-\frac{12p_{{\rm R}0}+(3/2)p_{{\rm N}0}}{\rho_0}
\frac{\omega}{u_{r0}}\frac{\partial \ln T_0}{\partial x}
-\frac{r_{{\rm s}0}-r_{{\nu}0}}{u_{r0}}
\frac{\partial \dot{q}}{\partial r} \right\} \nonumber\\
& &~~~~~~~~~\times\left\{xr_{{\rm s}0}\frac{r_{{\rm s}1}}{r_{{\rm s}0}}
+(1-x)r_{\nu 0}\frac{r_{\nu 1}}{r_{\nu 0}} \right\} \nonumber\\
& &-\frac{\dot{q}}{u_{r0}}\left(r_{{\rm s}0}\frac{r_{{\rm s}1}}{r_{{\rm s}0}}
-r_{{\nu}0}\frac{r_{{\nu}1}}{r_{{\nu}0}}\right)
-\frac{r_{{\rm s}0}-r_{{\nu}0}}{u_{r0}}
\frac{\partial \dot{q}}{\partial L_{\nu_{\rm e}}}L_{\nu_{\rm e} 0}
\frac{L_{\nu_{\rm e} 1}}{L_{\nu_{\rm e} 0}}
-\frac{r_{{\rm s}0}-r_{{\nu}0}}{u_{r0}}
\frac{\partial \dot{q}}{\partial r_{\nu}}r_{\nu 0}
\frac{r_{\nu 1}}{r_{\nu 0}}=0,
\label{eq20}
\end{eqnarray}
where, $p_{{\rm R}0}$, $p_{{\rm N}0}$ and $r_0$ are defined as

\begin{equation}
p_{{\rm R}0}\equiv\frac{11\pi^2}{180}\frac{k^4}{c^3 \hbar^3}T_0^4 ,
\label{eq21}
\end{equation}

\begin{equation}
p_{{\rm N}0}\equiv\frac{\rho_0 kT_0}{m_{\rm N}},
\label{eq22}
\end{equation}

\begin{equation}
r_0\equiv xr_{{\rm s}0}+(1-x)r_{{\nu}0}.
\label{eq23}
\end{equation}
All the variables, $u_{r1}/u_{r0}$, $\rho_1/\rho_0$, $T_1/T_0$,
$r_{{\rm s}1}/r_{{\rm s}0}$, $r_{{\nu}1}/r_{{\nu}0}$,
$L_{\nu_{\rm e} 1}/L_{\nu_{\rm e} 0}$ and an eigen-value $\omega$
can be complex.

The outer boundary conditions ($x=1$) become

\begin{equation}
\left(\frac{2}{r_{{\rm s}0}^2}\frac{\rho_{{\rm f}0}}{\rho_0}
\frac{u_{{\rm f}0}}{u_{r0}}+\frac{\omega}{r_{{\rm s}0}^{1/2}}
\frac{\rho_{{\rm f}0}}{\rho_0}-\rho_0 \omega r_{{\rm s}0}\right)
\frac{r_{{\rm s}1}}{r_{{\rm s}0}}
+\rho_0 u_{r0}\left(\frac{u_{r1}}{u_{r0}}+\frac{\rho_1}{\rho_0}\right)=0,
\label{eq24}
\end{equation}

\begin{eqnarray}
\lefteqn{\left(\frac{5}{2r_{{\rm s}0}^{5/2}}\frac{\rho_{{\rm f}0}}{\rho_0}
\frac{u_{{\rm f}0}^2}{u_{r0}^2}
+\frac{2\omega}{r_{{\rm s}0}}\frac{\rho_{{\rm f}0}}{\rho_0}
\frac{u_{{\rm f}0}}{u_{r0}}
-2\omega r_{{\rm s}0}\rho_0 u_{r0}\right)
\frac{r_{{\rm s}1}}{r_{{\rm s}0}}} \nonumber\\
& &+2\rho_0 u_{r0}^2 \frac{u_{r1}}{u_{r0}}
+(\rho_0 u_{r0}^2 +p_{{\rm N}0})\frac{\rho_1}{\rho_0}
+(4p_{{\rm R}0}+p_{{\rm N}0})\frac{T_1}{T_0}=0,
\label{eq25}
\end{eqnarray}

\begin{equation}
\left(\frac{1}{2r_{{\rm s}0}}\frac{u_{{\rm f}0}^2}{u_{r0}^2}
-\omega r_{{\rm s}0}^{1/2}\frac{u_{{\rm f}0}}{u_{r0}}
-\omega r_{{\rm s}0}u_{r0}\right)\frac{r_{{\rm s}1}}{r_{{\rm s}0}}
+u_{r0}^2 \frac{u_{r1}}{u_{r0}}
-4\frac{p_{{\rm R}0}}{\rho_0}\frac{\rho_1}{\rho_0}
+\frac{16p_{{\rm R}0}+5/2p_{{\rm N}0}}{\rho_0}\frac{T_1}{T_0}=0,
\label{eq26}
\end{equation}
where $\rho_{{\rm f}0}$, $u_{{\rm f}0}$ are density and velocity
of the free-fall at the shock in the unperturbed state.

Since we set the neutrino-sphere as the inner boundary ($x=0$)
and the density is fixed there by definition, we impose the condition

\begin{equation}
\frac{\rho_1}{\rho_0}=0,
\label{eq27a}
\end{equation}
at the inner boundary.
In addition, we assume

\begin{equation}
\frac{u_{r1}}{u_{r0}}=0,
\label{eq27b}
\end{equation}
at the inner boundary,
supposing that the infalling matter comes to rest on the neutron star.
Since the neutrino temperature is assumed to be unperturbed,
the variation of the luminosity is related with that of
the neutrino-sphere radius as,

\begin{equation}
\frac{L_{\nu_{\rm e} 1}}{L_{\nu_{\rm e} 0}}=2\frac{r_{{\nu}1}}{r_{{\nu}0}}.
\label{eq28}
\end{equation}

Above equations are solved as follows.
We first take trial values of $\omega$, $r_{{\nu}1}/r_{{\nu}0}$,
($r_{{\rm s}1}/r_{{\rm s}0}$ is set to be unity)
and solve the conditions (\ref{eq24}), (\ref{eq25}), (\ref{eq26}),
and equations (\ref{eq18}), (\ref{eq19}), (\ref{eq20})
from the outer boundary to the inner boundary.
The solution obtained this way does not, in general, meet the inner boundary
conditions (\ref{eq27a}), (\ref{eq27b}).
Then we improve the value of $\omega$, $r_{{\nu}1}/r_{{\nu}0}$
and repeat the procedure until (\ref{eq27a}), (\ref{eq27b})
are satisfied at the inner boundary.

We found for a given unperturbed solution
many modes with complex eigen-values
(probably an infinite number of overtones, in fact)
and only one mode with a real eigen-value.
All modes with complex eigen-values are damped,
namely, the real part of the eigen-value is negative,
owing to the thermal smearing effect of the neutrino irradiation.
We show only the results for the mode with a real eigen-value.
The growth rates are shown in figure \ref{fig3}.
We can see that the inner solutions are always stable,
while the outer solutions are always unstable.
Further, when the luminosity reaches the critical value,
both solutions become neutral.
The eigen-functions are shown in figure \ref{fig4}.
The ratios of the perturbation of luminosity to that of shock radius
are shown in figure \ref{fig5}.
Since the amplitude of luminosity perturbation is twice as much as
that of neutrino-sphere radius,
we can see that the perturbations of inner and outer boundaries
are in phase for the inner solutions,
whereas they are in opposite phase for the outer solutions.

We should note again that we investigated the stability only against radial
perturbations.
As several authors pointed out, spherical accretion shocks are,
in general, unstable against non-radial perturbations \citep{hou92,fog02,blo03}.
In this sense, we cannot say that the inner solutions are stable.
Detailed studies of the stability against non-radial perturbations
will be necessary with the effect of convection and
neutrino irradiation properly taken into account.

\section{The Effects of Rotation}

In this section, we discuss the effects of rotation on the revival
of shock.
We extend the one dimensional analysis in section 2
to the two dimensional case.
A brief summary of our results in this section has previously appeared
\citep{yam04}.
Since as we found in the previous section,
the radii of the outer solutions are unstable,
we consider only the inner solutions in this section.
The assumptions and simplifications mentioned in section 2
are also adopted here except for the assumptions
for the flows outside the shock, which will be mentioned later.
Assuming axial symmetry, we use the spherical coordinates.
Then the basic equations describing the steady flow
inside the shock are written as,

\begin{equation}
\frac{1}{r^2}\frac{\partial}{\partial r}(r^2 \rho u_r)+\frac{1}{r\sin\theta}
\frac{\partial}{\partial \theta}(\sin\theta\rho u_{\theta})=0,
\label{eq29}
\end{equation}

\begin{equation}
u_r\frac{\partial u_r}{\partial r}+\frac{u_{\theta}}{r}
\frac{\partial u_r}{\partial \theta}
-\frac{u_{\theta}^2 +u_{\phi}^2}{r}
=-\frac{1}{\rho}\frac{\partial p}{\partial r}-\frac{GM}{r^2},
\label{eq30}
\end{equation}

\begin{equation}
u_r\frac{\partial u_{\theta}}{\partial r}+\frac{u_{\theta}}{r}
\frac{\partial u_{\theta}}{\partial \theta}
+\frac{u_{r}u_{\theta}}{r}
-\frac{u_{\phi}^2\cot\theta}{r}
=-\frac{1}{\rho r}\frac{\partial p}{\partial \theta},
\label{eq31}
\end{equation}

\begin{equation}
u_r\frac{\partial u_{\phi}}{\partial r}+\frac{u_{\theta}}{r}
\frac{\partial u_{\phi}}{\partial \theta}
+\frac{u_{\phi}u_{r}}{r}
+\frac{u_{\theta}u_{\phi}\cot\theta}{r}=0,
\label{eq32}
\end{equation}

\begin{equation}
u_r\left(\frac{\partial\epsilon}{\partial r}-\frac{p}{\rho ^2}
\frac{\partial \rho}{\partial r}\right)
+\frac{u_{\theta}}{r}\left(\frac{\partial\epsilon}{\partial \theta}
-\frac{p}{\rho ^2}\frac{\partial \rho}{\partial \theta}\right)
=\dot{q},
\label{eq33}
\end{equation}
where ${\bf u}$ denotes velocity and the other notations are the same as
those in section 2.

Since we seek solutions with axial and equatorial symmetry,
we impose the following conditions,

\begin{equation}
u_{\phi}=u_{\theta}
=\frac{\partial u_r}{\partial \theta}
=\frac{\partial \rho}{\partial \theta}
=\frac{\partial T}{\partial \theta}=0,
\label{eq34}
\end{equation}
at $\theta=0$, and

\begin{equation}
u_{\theta}
=\frac{\partial u_r}{\partial \theta}
=\frac{\partial u_\phi}{\partial \theta}
=\frac{\partial \rho}{\partial \theta}
=\frac{\partial T}{\partial \theta}=0,
\label{eq35}
\end{equation}
at $\theta=\pi/2$.

The outer and inner boundaries are set at the shock surface
and neutrino-sphere, respectively.
The neutrino-sphere will be oblate due to centrifugal force
and, as a result, the neutrino flux will be anisotropic \citep{kot03}.
Such effect may have some influence on the explosion \citep{shi01}.
Since our concern here is the hydrodynamical effects of rotation
on the accretion flow,
we assume that the neutrino-sphere is spherical
and that the neutrino flux and temperature are isotropic.
Unlike in section 2, the shock surface is not spherical
and oblique to the flow, in general.

As in the one dimensional calculations,
we impose the condition that the density is $10^{11}$ ${\rm g/cm^3}$,
at the inner boundary.
The Rankine-Hugoniot relations for oblique shocks are imposed
at the outer boundary.
As for the flows outside the shock,
we take the following approximations:
i) the flows are radial except for rotation and free fall.
ii) the density is independent of latitude,
thus the accretion isotropic outside the shock.
iii) the rotation frequency is independent of latitude 
and the specific angular momentum is conserved along each stream curve.
Although these assumptions are artificial and not self-consistent,
since the centrifugal force is small compared with
the gravity or the inertia, the radial component of the velocity
must be predominant in reality.

We define the parameter $f$ as the rotational frequency
at the radius of $1000$ km.
We performed the calculations for $f=0.03$ ${\rm s}^{-1}$
and $0.1$ ${\rm s}^{-1}$,
which correspond to the specific angular momenta
averaged over the whole sphere outside the shock wave surface of $1.256$
and $4.187\times 10^{15} {\rm cm^2/s}$, respectively.
(The definition of the specific angular momentum is employed
by \citet{heg00,heg03} in their calculations of stellar evolutions).

In figure \ref{fig6}, we show the stream curves for the parameters
$\dot{M}=2.0$ $M_{\odot}/{\rm s}$,
$L_{\nu_{\rm e}}=7.0 \cdot 10^{52}$ ${\rm ergs/s}$,
and $f=0.1$ ${\rm s}^{-1}$.
We can see that the flow is pushed toward the equatorial plane owing to
centrifugal force.
The stream curves bend rather abruptly near the neutrino-sphere
because the infalling matter is decelerated as it lands
onto the neutron star surface
and the radial velocity falls rapidly there.
It is also noted that the shock surface is slightly
deformed into an oblate configuration,
and the flows are directed toward the rotational axis at large radii.

The angular dependence of the radial velocity inside the shock is shown
in figure \ref{fig7}.
At large radii, the velocity is larger near the rotation axis than
the equatorial plane, whereas the opposite is true near the inner boundary.
This behavior is attributed to the variation of cross section of the flow.
The theory of Laval nozzle tells us that subsonic flows are decelerated
when the cross section of nozzle gets larger, vice versa \citep{lan87}.
As we saw in figure \ref{fig6}, the flow is bent toward the equatorial plane
near the neutrino-sphere but toward the rotational axis near the shock.
As a result, at small radii,
the cross section of flow becomes larger near the axis,
whereas it gets smaller near the equatorial plane,
and vice versa at large radii.
Thus, we obtain the observed angular dependence of velocity.

Next we discuss the effect of rotation on the critical luminosity.
In figure \ref{fig8}, we show the dependence of the critical luminosity
on the mass accretion rate.
We can see that the smaller the mass accretion rate is,
or the larger the rotation frequency is,
the more reduced the critical luminosity is.
According to the results of our calculations,
if $f$ is 0.03 ${\rm s}^{-1}$, and the accretion rate is 0.1 $M_{\odot}$/s,
the critical value is about $25\%$ smaller than
that of the spherically symmetric flow with the same mass accretion rate.
If the rotation parameter is 0.1 ${\rm s}^{-1}$ the effect is more remarkable.
For the mass accretion rates of 1.0 $M_{\odot}$/s and 0.5 $M_{\odot}$/s,
we get the critical luminosities reduced by about $25$ and $43\%$ respectively,
from the value for the spherically symmetric flow.

Finally, we discuss the fate of the shock wave
when the critical luminosity is reached.
In section 2, we discussed the movement of the steady shock
for the inner solutions;
the radial derivative of the infall velocity increases
when the neutrino luminosity is raised,
and it becomes nearly equal to that of $u_{\rm post}(r)$
when the luminosity reaches the critical value.
The angular dependence of the velocity
at the critical luminosity is shown in figure \ref{fig7}.
We can see that the derivative is largest at $\theta=0$.
This suggests that the steady flow cease to exist first at the axis.
Thus, we can infer that the stalled shock is revived and start to
propagate outwards first at the rotational axis and
that the jet-like explosion is followed,
although time-dependent numerical simulations are necessary to confirm this.

This break up of shock at the rotational axis and
the resulting reduction of the critical luminosity are also attributed
to the changes of radial flow velocities, which can be understood based 
on the theory of Laval nozzle, as we saw above. The variation of 
flow-cross-section causes the acceleration of the radial flow at large radii 
near the rotation axis whereas it leads to the deceleration near the 
equatorial plane. As a result, the revival of the stalled shock is more easily 
attained at the rotational axis for lower neutrino luminosities 
than in the spherically symmetric flows.

\section{Summary and Discussion}

In this paper, we investigated the behavior of the stalled shock
in the supernova core and the effects of rotation on its revival.
The spherically symmetric accretion flows were re-investigated first.
We found that there exist two branches of solutions when the luminosity
of the irradiating neutrinos is below the critical value.
The steady solution does not exist when the luminosity is higher than
the critical value.

\begin{figure}
\plotone{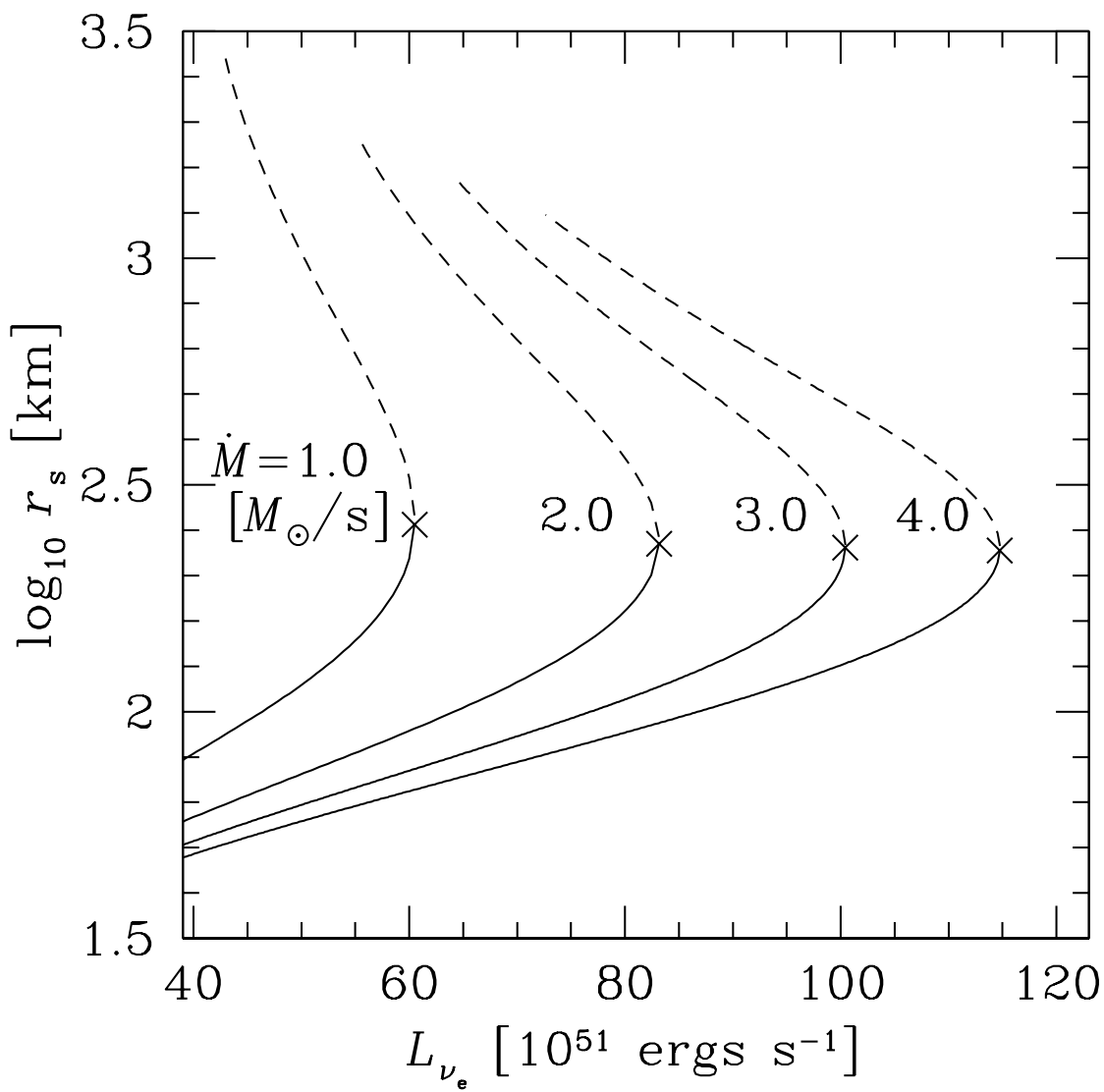}
\caption{Radii of shock surface.
Solid curves denote the inner solutions
for $\dot{\rm M}=1.0, 2.0, 3.0,
4.0$ ${\rm M}_{\odot}/{\rm s}$ from left to right.
Dashed curves display the outer solutions
for $\dot{\rm M}=1.0, 2.0, 3.0,
4.0$ ${\rm M}_{\odot}/{\rm s}$ from left to right.
The crosses represent the critical points.
\label{fig2}}
\end{figure}

\begin{figure}
\plotone{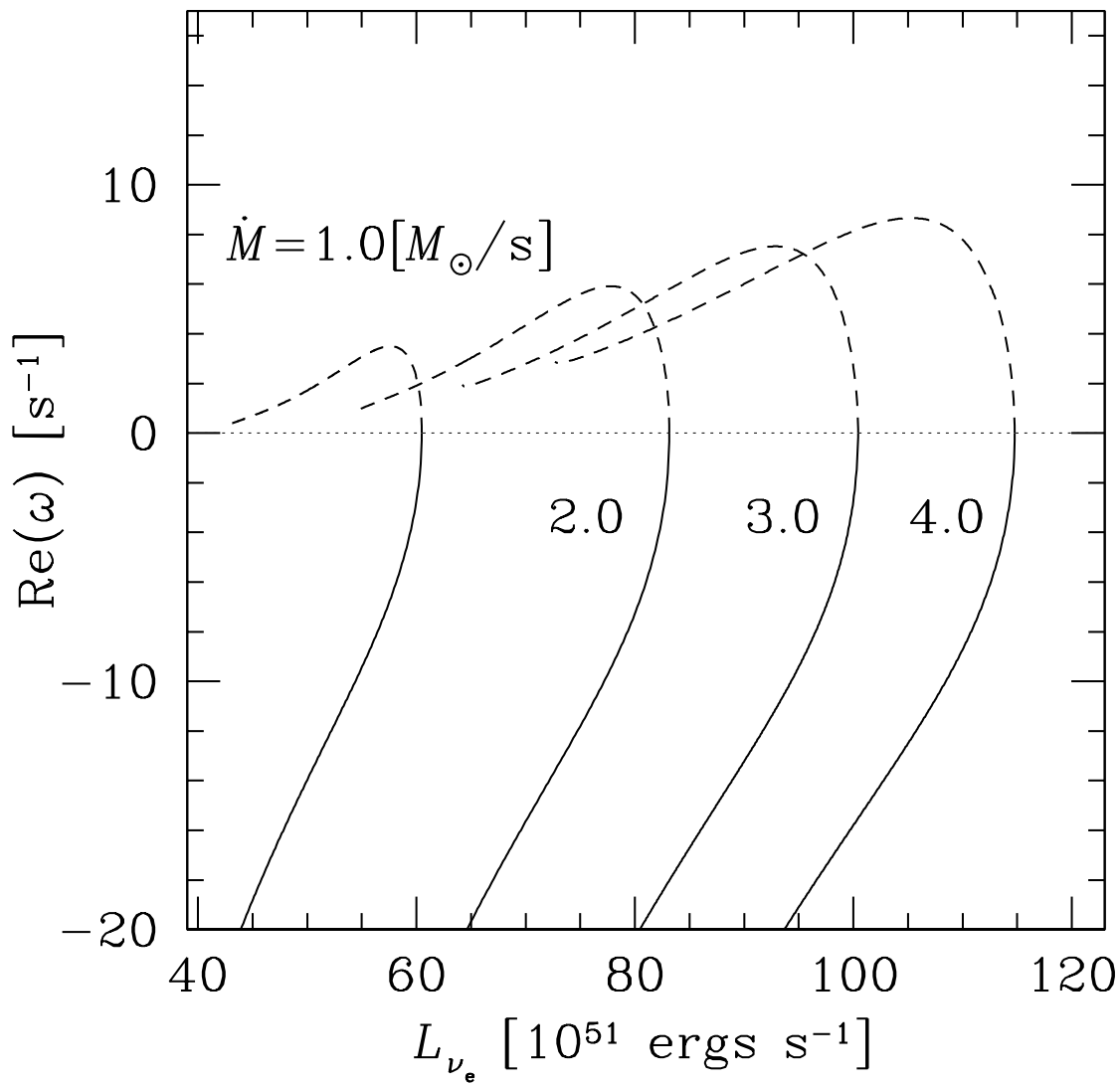}
\caption{Growth rates for radial perturbations.
Solid curves show the inner solutions
for $\dot{\rm M}=1.0, 2.0, 3.0,
4.0$ ${\rm M}_{\odot}/{\rm s}$ from left to right.
Dashed curves depict the outer solutions
for $\dot{\rm M}=1.0, 2.0, 3.0,
4.0$ ${\rm M}_{\odot}/{\rm s}$ from left to right.
\label{fig3}}
\end{figure}

\begin{figure}
\plottwo{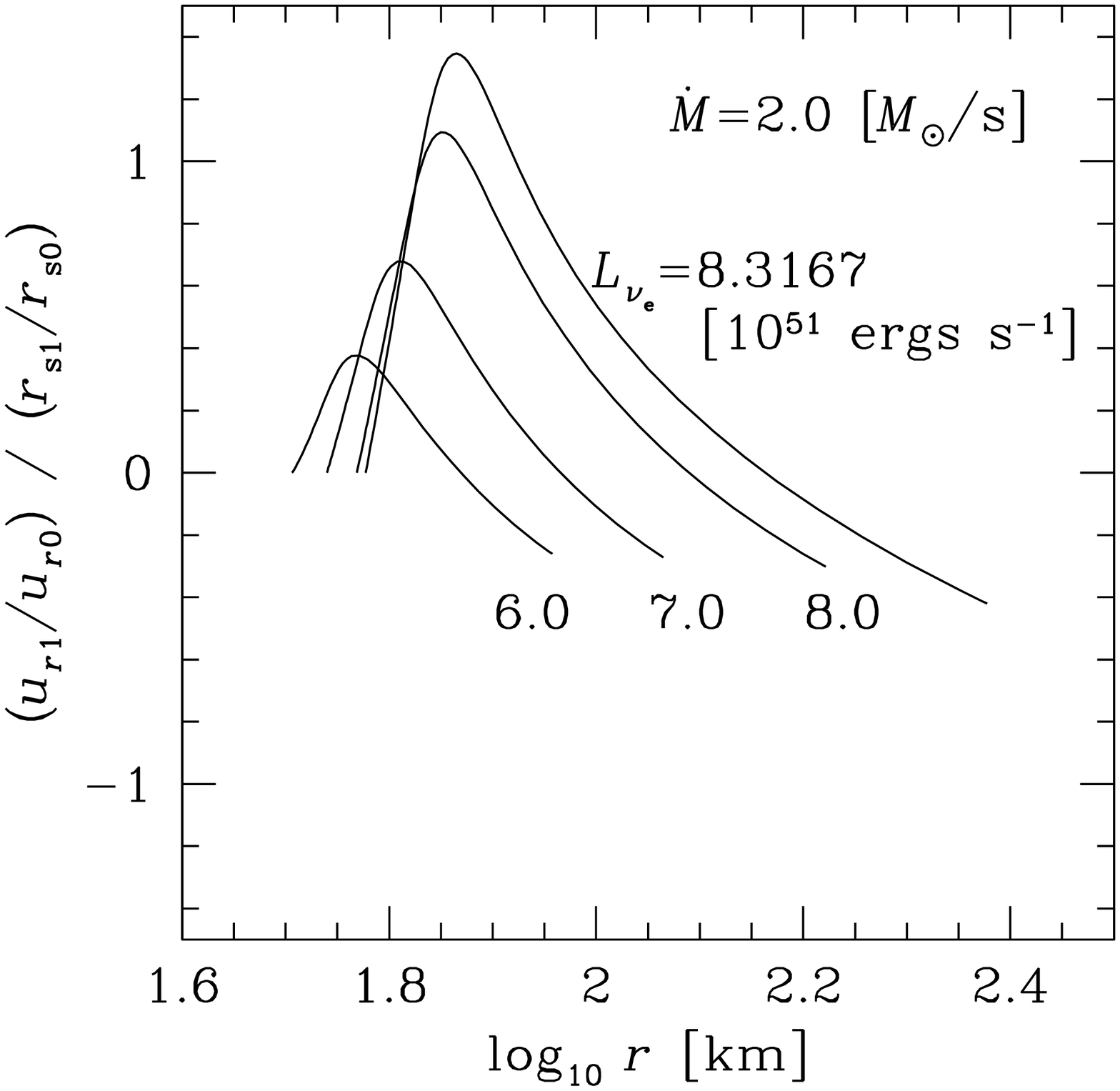}{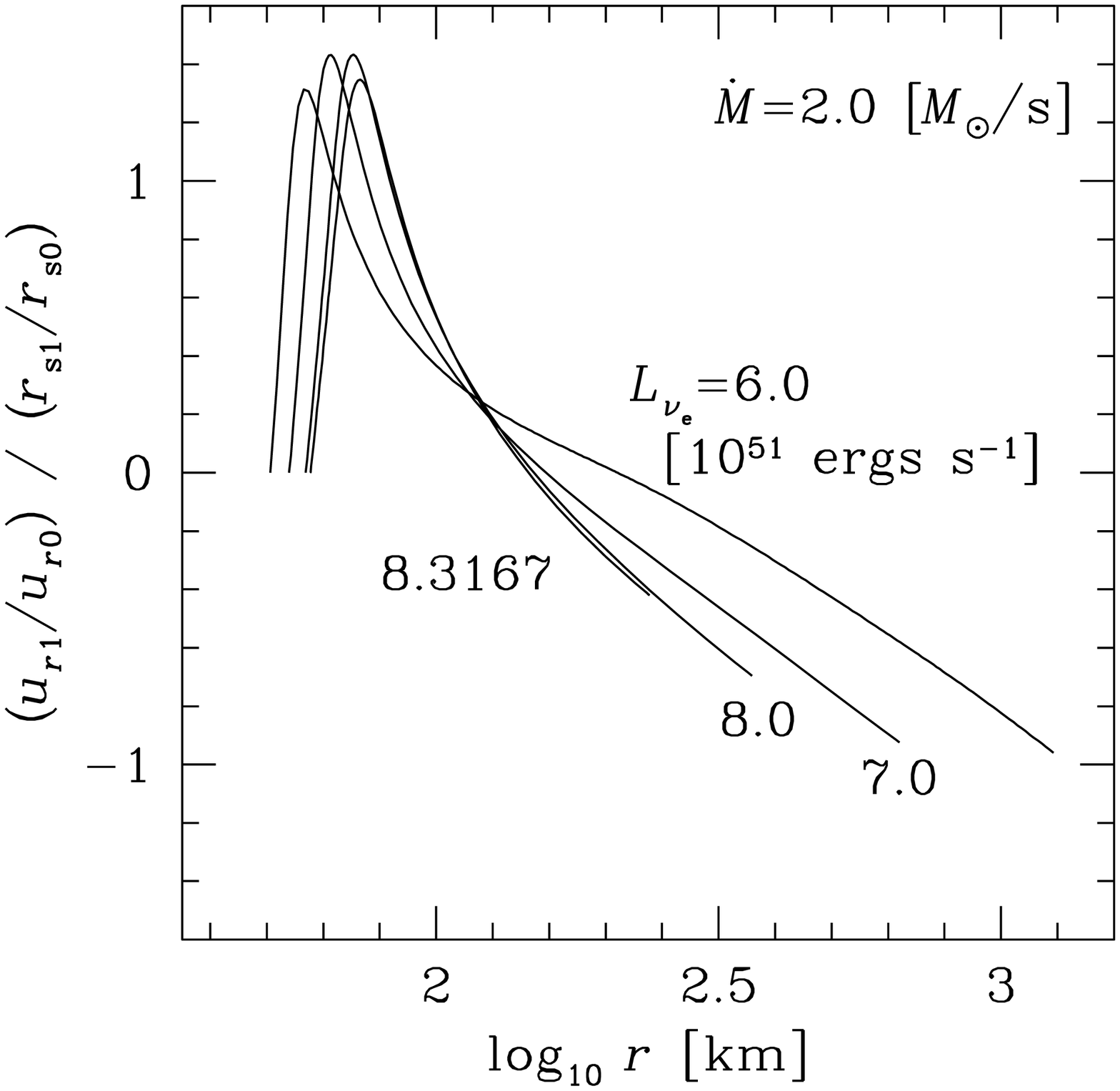}
\caption{a. eigen-functions of radial perturbations
for the inner solutions with $\dot{\rm M}=2.0$ ${\rm M}_{\odot}/{\rm s}$,
${\rm L}_{\nu_{\rm e}}=6,7,8\cdot 10^{52}$ ergs/s
and $8.3167 \cdot 10^{52}$ ergs/s
(critical value) from left to right.
b. eigen-functions of radial perturbations
for the outer solutions with $\dot{\rm M}=0.2$ ${\rm M}_{\odot}/{\rm s}$,
${\rm L}_{\nu_{\rm e}}=6,7,8\cdot 10^{52}$ ergs/s
and $8.3167 \cdot 10^{52}$ ergs/s from right to left near the shock.
Note that the functions are defined
in different ranges of radius for different models.
\label{fig4}}
\end{figure}

\begin{figure}
\plotone{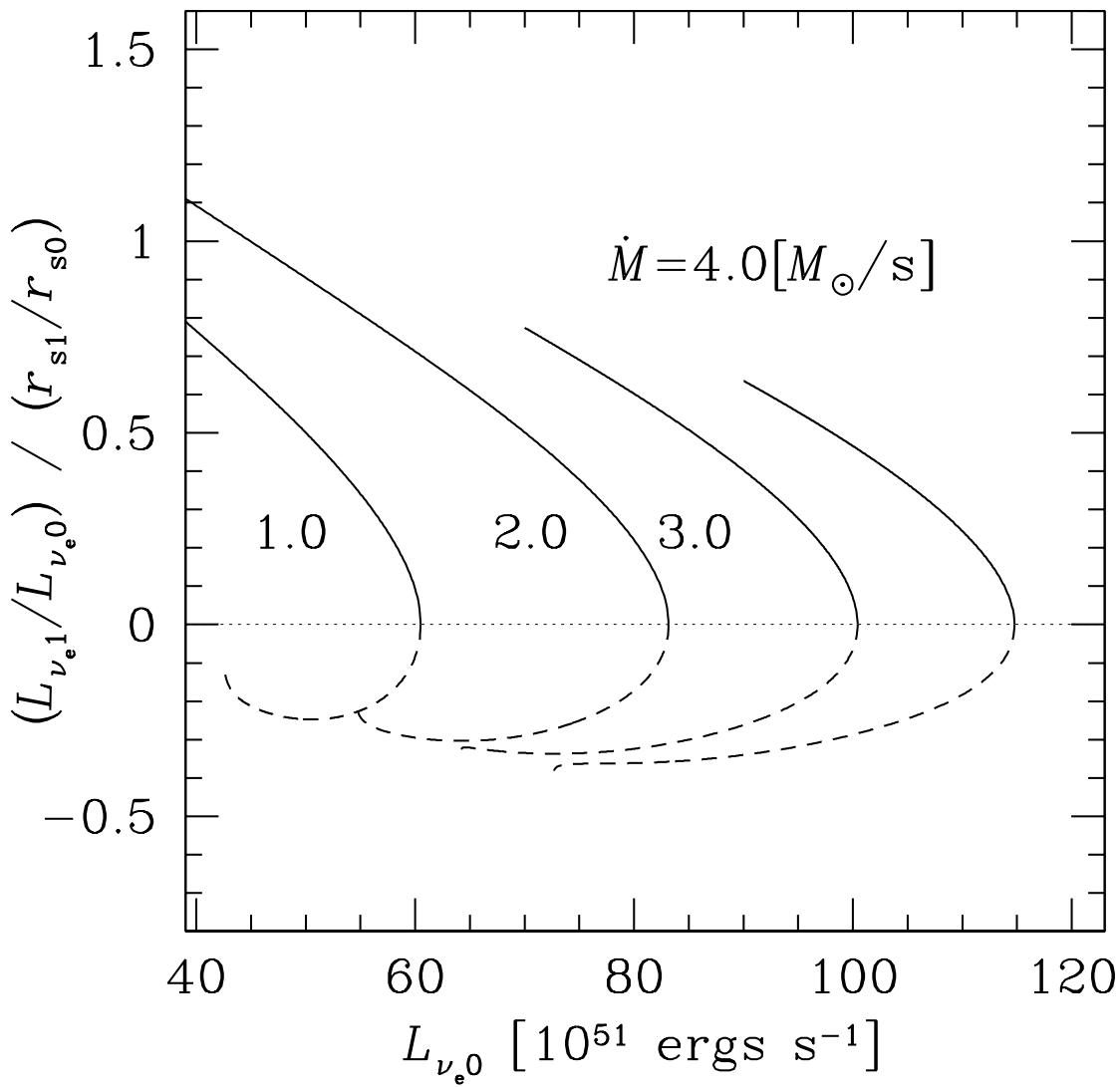}
\caption{Ratio of perturbation of luminosity to that of shock radius.
Solid curves denote the inner solutions
for $\dot{\rm M}=1.0, 2.0, 3.0,
4.0$ ${\rm M}_{\odot}/{\rm s}$ from left to right.
Dashed curves present the outer solutions
for $\dot{\rm M}=1.0, 2.0, 3.0,
4.0$ ${\rm M}_{\odot}/{\rm s}$ from left to right.
\label{fig5}}
\end{figure}

\begin{figure}
\plotone{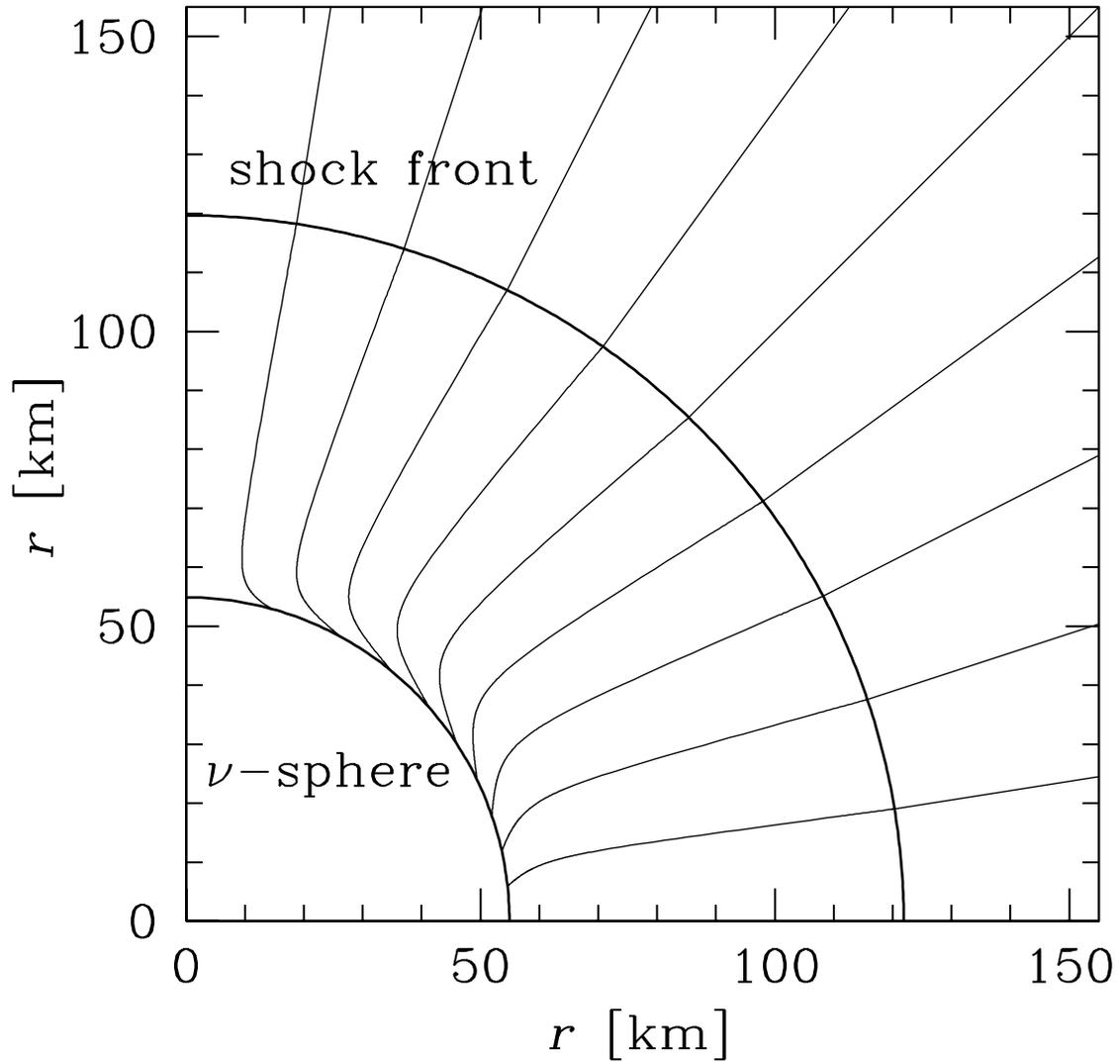}
\caption{Meridian section of a collapsing star and stream curves
for $\dot{\rm M}=2.0$ ${\rm M}_{\odot}/{\rm s}$,
${\rm L}_{\nu_{\rm e}}=7\cdot 10^{52}$ ergs/s
and rotation frequency 0.1 ${\rm s}^{-1}$ at ${\rm r}=1000$ km.
\label{fig6}}
\end{figure}

\begin{figure}
\plotone{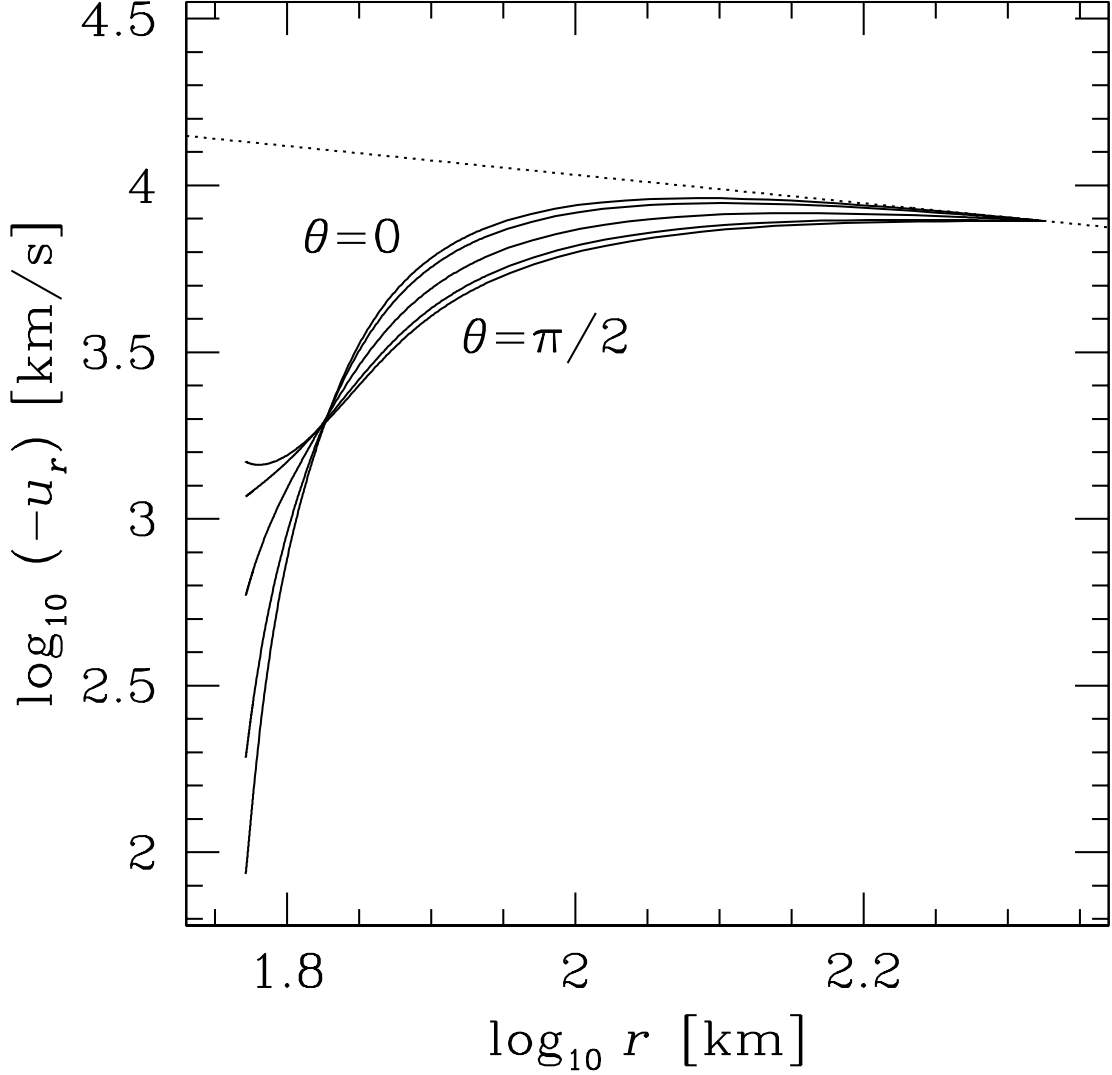}
\caption{Radial velocities at latitude
$\theta=0,\pi/8,\pi/4,3\pi/8$ and $\pi/2$
for the rotating steady accretion flow
with $\dot{\rm M}=2.0$ ${\rm M}_{\odot}/{\rm s}$,
${\rm L}_{\nu_{\rm e}}=8.10 \cdot 10^{52}$ ergs/s (critical value)
and rotation frequency 0.1 ${\rm s}^{-1}$ at ${\rm r}=1000$ km
from top to bottom.
Dotted line shows the downstream value
satisfying the Rankine-Hugoniot relations at each radius.
\label{fig7}}
\end{figure}

\begin{figure}
\plotone{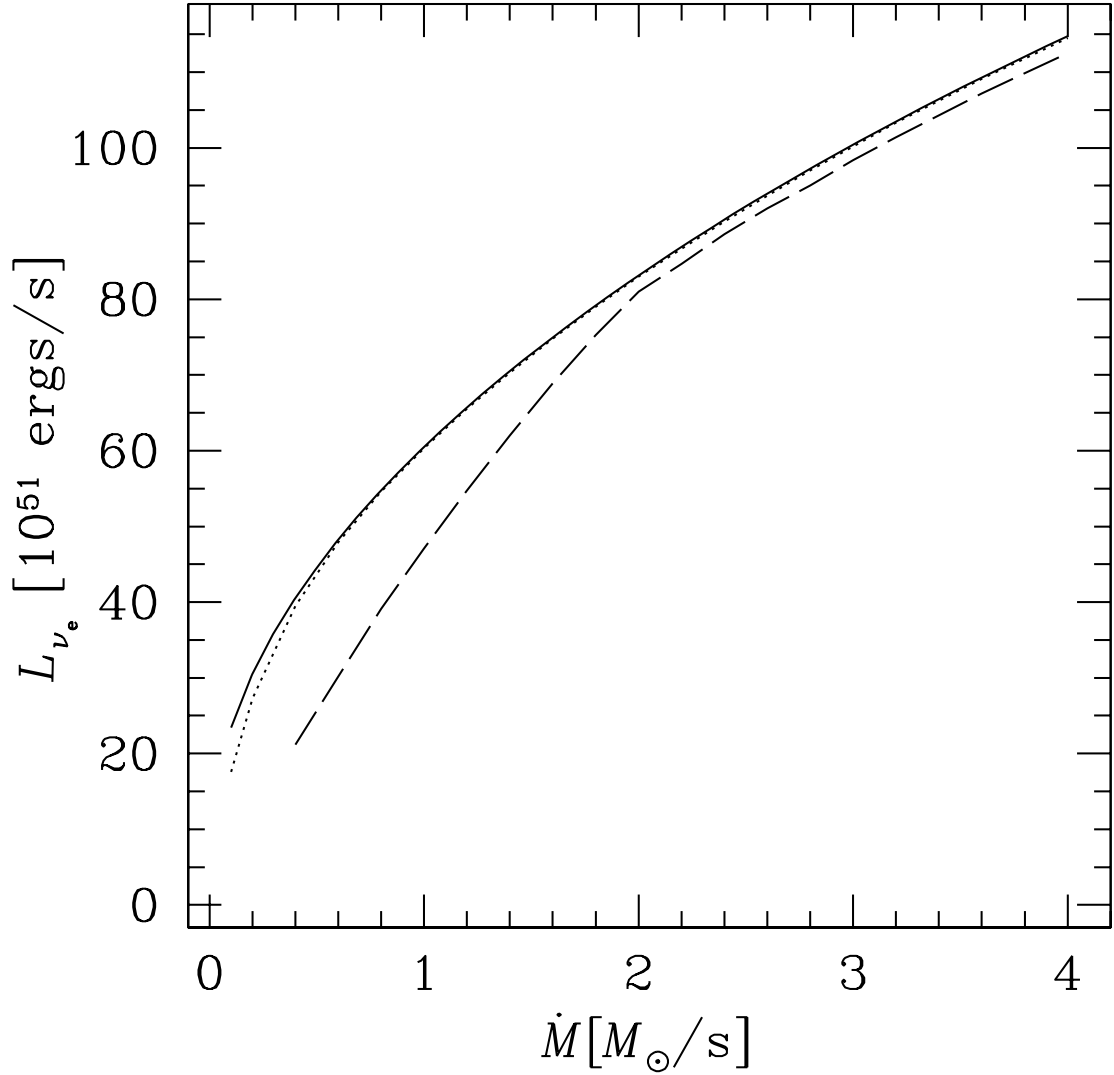}
\caption{Critical luminosity.
Solid curve denotes the spherically symmetric case,
and the others correspond to the rotational models.
The rotation frequencies are 0.03, 0.1 ${\rm s}^{-1}$ at 1000 km
for dotted and dashed curves, respectively.
No steady solution exists for the luminosity larger than the critical value.
\label{fig8}}
\end{figure}

In order to consider which solution is realized below the critical luminosity,
we then examined the stability of the spherical accretion flows
against radial perturbations,
and found that the inner solutions are stable while outer solutions
are unstable.
In fact, for small luminosities expected for the early phase of the shock
stagnation, the radii of the outer solutions
are extremely large (figure \ref{fig2}).
Thus the outer solutions will not be realized in reality.
The inner solutions are always stable until
the neutrino luminosity reaches the critical value,
where the two branches merge and the solution becomes neutrally stable.
It should be noted, however, that the stability of the outer solutions may be 
affected by rotation, which will be a future work.

We did not discuss the stability against non-radial perturbations,
since our model are convectively unstable.
Several authors showed the spherical accretion shock is unstable
against non-radial perturbations.
\citet{blo03} demonstrated the instability for adiabatic perturbations
by numerical simulations.
\citet{hou92} examined such a stability taking into account
the cooling processes whose rates are determined locally.
In the present situation, however, the heating rates are not
determined locally because the flow is irradiated by neutrinos coming out of
the proto-neutron star;
the perturbations of the flow affect
the neutrino luminosity, and the latter then affects the former,
by varying the heating and cooling rates.
Thus, the problem is different from those they treated.
The detailed analysis in such a situation is a future work.
Furthermore, the stability analysis with rotation taken into account
is also an interesting issue.

Finally, we discussed the effects of rotation on the revival of the stalled
shock.
We showed that rotation lowers the critical luminosity.
However, we claimed that the shock revival
would first take place at the rotation axis.
We took the rotation frequency of 0.03 and 0.1 ${\rm s}^{-1}$
at the radius of 1000 km and found that the rotation effect is substantial.
The rotation frequency of newly born neutron stars
guessed from the spin down rate of young pulsars
is about several ten per second \citep{kas94,swa01}.
Thus the frequency we adopted here is one order of magnitude higher than
the value inferred from the observations of young pulsars.
It might be possible, however, that this discrepancy is removed by some
mechanisms such as magnetic breaking in the late phase of explosion,
which could take away the excessive angular momentum
from the proto-neutron stars.
In fact, the values we adopted are consistent with
those obtained in calculations of the evolution
of rotating stars \citep{heg03}.
Anyway, the angular momentum distribution inside massive stars
are not yet well understood
because of our poor knowledge on the mechanism of angular momentum transform
(cf. \citet{heg00,heg03}).
If the angular momentum of the infalling matter is large enough,
our results suggest that rotation assists the shock revival.

In this paper, we employed various assumptions and simplifications
in the formulations.
We assumed that the neutrino-sphere is spherical
and the neutrino flux is isotropic.
If the collapsing star is rotating,
the proto-neutron star will have angular momentum.
\citet{kot03} investigated how the neutrino-sphere is affected by rotation.
They showed that the neutron star becomes oblate and, as a result,
the emerging flux is more concentrated to the rotation axis.
One the other hand, \citet{shi01} studied how the explosion is affected when
the neutrino flux is greater near the rotation axis,
and claimed that such neutrino anisotropy can trigger a jet-like explosion
in otherwise failed models.
Such effects were neglected in this paper just for simplicity and
are expected to further reduce the critical luminosity,
which we will study in the future.

The convection should be also affected by rotation.
\citet{fry00} studied such effects by numerical simulations.
They found that the centrifugal force reduce the effective gravity, and,
thus, the efficiency of convection is lowered near the equatorial plane.
As a result, the explosion energy was also decreased.
The explosion, however, becomes jet-like.
In order to investigate the effect of convection
in the time-independent models,
it is necessary to introduce some simplification,
such as a mixing length theory or other formulation to smooth the entropy
gradient.
This is currently being undertaken.

Finally the collapsing stars may have magnetic fields, which are then
amplified by rotation.
The recent works suggest that the magnetic fields are amplified by
the differential rotation or magneto-rotational instability
\citep{aki03,kot04,ard04}.
If the magnetic energy is amplified to be
comparable to the rotational energy, it must affect the accretion flow
and the behavior of the stalled shock as they suggested.
This issue is also a future work.

\acknowledgments

\clearpage

\end{document}